\documentclass[12pt,a4paper]{article}
\usepackage{graphicx}
\usepackage{times}
\title{\bf Dual Active Galactic Nuclei in Nearby Galaxies\footnote{}}
\author{M.~Das$^1$\thanks{mousumi@iiap.res.in}, K.~Rubinur$^{1,2}$, P.~Kharb$^3$,\\
A.~Varghese$^4$, K.~Navyasree$^4$, and A.~James$^4$\\
\vspace{1cm}\\
\normalsize $^1$ Indian Institute of Astrophysics, Bangalore 560034, India\\ 
\normalsize $^2$ Pondicherry University, 605014 Pondicherry, India\\
\normalsize $^3$ National Center for Radio Astrophysics-Tata Institute of Fundamental Research, Pune 411007, India\\
\normalsize $^4$ Christ University, Hosur Road, Bengaluru 560029, India}
\date{\mbox{}}
\begin{document}
\maketitle
\pagestyle{empty}
%
%
\def\bull{\vrule height .9ex width .8ex depth -.1ex}
\makeatletter
\def\ps@plain{\let\@mkboth\gobbletwo
\def\@oddhead{}\def\@oddfoot{\hfil\scriptsize\bull\quad
"First Belgo-Indian Network for Astronomy \& astrophysics (BINA) workshop'', held in Nainital (India), 15-18 November 2016 \quad\bull}%
\def\@evenhead{}\let\@evenfoot\@oddfoot}
\makeatother
%
%
\def\beginrefer{\section*{References}%
\begin{quotation}\mbox{}\par}
\def\refer#1\par{{\setlength{\parindent}{-\leftmargin}\indent#1\par}}
\def\endrefer{\end{quotation}}
%
%
{\noindent\small{\bf Abstract:} 
Galaxy mergers play a crucial role in the
formation of massive galaxies and the buildup of their bulges. An important aspect of the merging
process is the in-spiral of the supermassive black-holes (SMBHs) to the centre of the
merger remnant and the eventual formation of a SMBH binary. If both the SMBHs
are accreting they will form a dual or binary active galactic nucleus (DAGN). The final merger remnant
is usually very bright and shows enhanced star formation. In this paper we summarise the 
current sample of DAGN from previous studies and describe methods that can be used to identify
strong DAGN candidates from optical and spectroscopic surveys. These methods
depend on the Doppler separation of the double peaked AGN emission lines, the
nuclear velocity dispersion of the galaxies and their optical/UV colours. We describe two
high resolution, radio observations of DAGN candidates that have been selected based on their double
peaked optical emission lines (DPAGN). We also examine whether DAGN host galaxies have higher star 
formation rates (SFRs) compared to merging galaxies that do not appear to have DAGN. We find that the 
SFR is not higher for DAGN host galaxies. This suggests that the SFRs in DAGN host galaxies is due to the 
merging process itself and not related to the presence of two AGN in the system.     
}
%
%
\section{Introduction}

Galaxy mergers are the most powerful drivers of galaxy evolution in our low redshift Universe. During 
these mergers the nuclear supermassive black holes (SMBHs) of the individual galaxies sink into the 
center of the merger remnant (Begelman, Blandford \& Rees 1980). When the SMBH pair members are at separations 
of $<$~10~kpc, they are referred to as 
dual SMBHs and when they are at separations $<$~100~pc, they are called binary SMBHs (Burke-Spolaor et al. 2014). 
If both the SMBHs are accreting mass they can form active galactic nuclei (AGN) pairs.
Depending upon their AGN separations they are called dual or binary AGN and can be detected at optical 
(Liu et al. 2010), radio (e.g. Rodriguez et al. 2006) and X-ray wavelengths 
(e.g. Koss et al. 2012; Comerford et al. 2013). Radio jets may arise from one 
or both AGN (Rees et al. 1982). The SMBHs will eventually become gravitationally bound at separations 
of a few parsecs and may be enveloped by a common circumbinary disk (Hayasaki, Mineshige \& Ho 2008). They will 
finally coalesce releasing enormous amounts of gravitational radiation (Abbott et al. 2016). 

From simulations it is well understood that dual nuclei
play an important role in driving gas towards the centers of merger remnants and triggering
star formation (Mayer et el. 2007) leading to the merger of the SMBHs (Kelley et al. 2017).
This is clearly seen in observations of gas rich mergers such as ultraluminous Infrared Galaxies (ULIRGs)
where the merging is accompanied by enormous amounts of star formation (Sanders et al. 1988). The
dense gas collects not only around the individual nuclei but also between
the nuclei, resulting in a velocity gradient in the molecular gas distribution; such
a morphology is expected from gas dynamics. However, if the nuclei are accreting (i.e. are AGN)
or associated with intense star formation, the AGN/starburst activity will also be accompanied
by galactic scale outflows of hot, ionized gas that can drive out the cold molecular gas from the
nucleus and quench star formation (Dasyra et al. 2014; Garcia-Burillo et al. 2015). This negative
feedback of AGN or starburst activity can help regulate the growth of the galaxy bulge as well as
the SMBHs (e.g Silk \& Rees 1998). Thus in merging, luminous starburst galaxies, there are two
competing mechanisms - the two cores driving the gas inwards and the outflows driving the nuclear
gas outwards (Sakamoto et al. 2014). 

In this paper we mainly focus on two aspects of dual AGN (DAGN) in merger remnants - their detection and 
the star formation associated with the DAGN. In the following section we describe the 
observational methods of detecting DAGN and the current confirmed DAGN sample. We then discuss radio 
observations of two double peaked emission line nuclei, KISSR~1494 and 2MASX~J1203. In the remaining two
sections we estimate the effect of AGN pairs on star formation in gas rich merger remnants and then
summarise our results.  
 
\section{DAGN in Galaxies : their detection and the current known sample}

There are several methods by which DAGN can be detected in galaxies and we have discussed this in Rubinur
et al. (2017) but we briefly describe it here as well. The earliest DAGN were discovered by chance through 
the optical variability studies of quasars. The most famous example is the quasar OJ~287 whose optical 
AGN variability has been tracked for several decades now (Sillanpaa et al. 1988; Lehto \& Valtonen 1996). 
The variability arises due to the time dependent variation of the SMBH separation. This will affect the 
mass accretion rate of the individaul AGN and hence produce a periodicity in the emisison from the DAGN,
as has been shown in various models (e.g. Valtonen et al. 2008). There are now several programs that track optical 
variability of quasars and this may lead to new discoveries.

Another indirect method that has led to DAGN detections is the observation of S or X-shaped radio jets in 
the centers of radio bright galaxies (Begelman, Blandford \& Rees 1980). The S-shaped radio jet morphologies
are thought to arise if there is a companion black hole orbiting an AGN. The periodic pertubations caused
by the companion can cause the accretion disk to precess and result in precessing radio jets that will 
appear as S-shaped radio jets (Rubinur et al. 2017). The X-shaped radio sources, however, may arise due  
to the spin flip of one of the SMBHs in the DAGN, a very good example is NGC~326 (Murgia et al. 2001;
Hodges-Kluck \& Reynolds 2012). There are other explanations 
for S or X-shaped radio jets such warping of the accretion disks (Pringle 1997) or back flowing gas
(Leahy \& Williams 1984).

In the past two decades after the advent of large spectroscopic surveys, double peaked emission lines have been 
detected in the optical spectra of AGN and the galaxies are called double peaked AGN (DPAGN; Zhou et al. 2004). 
The Doppler shift between the two line components can be due to several reasons apart from two AGN : rotating 
nuclear disks , AGN outflows or collimated jets (Kharb et al. 2015 and references therein). For example, Figure~1  
shows the double peaked emission 
lines in [OIII]. Thus to confirm the presence of two AGN in a DPAGN high resolution 
radio observations (e.g. Tingay \& Wayth 2011; Kharb et al. 2015; Rubinur et al. 2017) or X-ray observations are
essential (e.g. Liu et al. 2013; Comerford et al. 2015). To date, about 30\% of all confirmed DAGN are from 
DPAGN samples (McGurk et al. 2015) and hence this may be a good technique, though high resolution 
imaging is essential. On top of the double peaked lines, a further discriminator could be choosing DPAGN nuclei that
have high stellar velocity dispersion ($\sigma$) (Rubinur et al. 2017, in preparation). This is because a pair of 
SMBHs in the nuclear region will dynamically heat the stars (Merritt \& Milosavljevic 2005).   

There are several reasons why detecting two AGN in the centers of galaxies is important. In the light of recent 
detection of gravitatational radiation, it is important to determine a sample of closely interacting SMBHs whose 
gravitational radiation can be detected using pular timing arrays or $\mu$Hz detectors such as eLISA. Such SBMH 
pairs can be detected only through their emission in the electromagnetic spectrum which occurs when they are 
accreting as AGN. Apart from their interaction, dual or binary AGN will drive the cold gas into the centers of 
galaxies resulting in nuclear star formation, strong winds, outflows and AGN feedback. Thus dual AGN also play an 
important role in galaxy evolution, especially in gas rich merger remnants. 
  
\begin{table}
\vspace{-1cm}
\caption{The table shows the sample of confirmed DAGN that have been
collected from the literature along with the reference.}
\small
\begin{center}
\begin{tabular}{| l | r r r r r|}
\hline
No & Galaxy Name  & RA & Declination & Redshift & Reference \\
\hline

1  & LBQS 0103-2753  & 01h05m34.7s    & -27d36m59s    & 0.85     &    Junkkarinen et al. (2001)    \\
2  & NGC 6240         & 16h52m58.9s    & +02d24m03s    & 0.02     &    Komossa et al. (2003)\\
3  & 4c + 37.11       & 04h05m49.2s    & +38d03m32s    & 0.06     &   Rodriguez et al. (2006) \\   
4  & 3c75             & 02h57m41.6s    & +06d01m29s    & 0.02     &   Hudson et al. (2006) \\
5  & MRK 463          & 13h56m02.9s    & +18d22m19s    & 0.05     &    Bianchi et al. (2008)\\
6  & CID 42          & 14h47m06.6s    & +11d35m29s    & 0.03     &      Civano et al. (2010) \\
7  & 2MASX J11312609-0204593   & 11h31m26.1s  & -02d04m59s & 0.15 &   Liu et al. (2010)\\   
8  & SDSS J133226.34+060627.3  & 13h32m26.3s & +06d06m27s & 0.21  &   Liu et al. (2010)\\  
9  & NGC 326          & 00h58m22.7S    & +26d51m55s    & 0.05     &    Murgia et al. (2001) \\
10 & NGC 3393         & 10h48m23.4s    & -25d09m43s    & 0.01     &    Fabbiano et al. (2011)\\
11 & 2MASX J11085103+0659014 & 11h08m51.0s  & +06d59m01s & 0.18  &    Liu et al. (2010)\\
12 & SDSS J150243.1+111557    & 15h02m43.1s    & +11d15m57s    &  0.39   & Fu et al. (2011)\\
13 & SDSS J095207.62+25527.2 & 09h52m07.6s & +25d52m57s & 0.34   & McGurk et al. (2011) \\
14 & MRK 739          & 11h36m29.1s    & +21d35m46s    & 0.03     &    Koss et al. (2012)\\
15 & IRAS 05589+2828  & 06h02m10.7s    & +28d28m22s    & 0.03     &    Koss et al. (2012)\\
16 & SDSS J142607.71+353351.3  & 14h26m07.7s & +35d33m51s & 1.16  & Barrows et al. (2012)\\
17 & MRK 266                  & 13h38m17.5s    & +48d16m37s    &  0.03 &    Mazzarella et al. (2012)\\
18 & ESO 509-IG066 NED 02     & 13h34m40.8s    & -23d26m45s    &  0.03 &    Koss et al. (2012)\\
19 & IRAS 03219+4031          & 03h25m12.7s    & +40d41m58s    & ----  &   Koss et al. (2012)\\
20 & NGC 3227                 & 10h23m30.6s    & +19d51m54s    &  0.00   &    Koss et al. (2012)\\
21 & J171544.02+600835.4      & 17h15m44.038s  & +60d08m35.29s &  0.16   & Comerford et al. (2013)\\                           
22 & SDSS J102325+324348    & 10h23m25.6s    &+32d43m49s  & 0.13  & Muller-Sanchez et al. (2015)    \\
23 & SDSS J115822+323102    & 11h58m22.6s    & +32d31m02s & 0.17  & Muller-Sanchez et al. (2015)\\
24 & SDSS J162345+080851    & 16h23m45.2s    & +08d08m51s & 0.20  & Muller-Sanchez et al. (2015)\\
25 & SDSS 114642.47+511029.6  & 11h46m42.5s    & +51d10m30s    &  0.13 & McGurk et al. (2015)\\
26 & SDSS J112659.59+294442.8 & 11h26m59.538s  & +29d44m42.78s &  0.10 & Comerford et al. (2015)\\

\hline 
\end{tabular} 
\end{center} 
\end{table}

\section{Radio Observations of the DPAGN galaxies KISSR~1494 and 2MASX~J1203}

\subsection{KISSR~1494}

This galaxy hosts a Seyfert~2 nucleus which shows double peaked emission lines in its Sloan Digital Sky Survey 
(SDSS) spectrum (for details and the spectrum of KISSR~1494 see Kharb et al. 2015). The H$\alpha$ and H$\beta$ lines in KISSR~1494 have equal peaks; such DPAGN are often referred to as an EPAGN. The equal peaks could be due to a nuclear disk or two AGN
(Smith et al. 2012). We conducted high resolution, Very Long Baseline Interferometry (VLBI) radio observations of 
KISSR~1494 using the Very Long Baseline Array (VLBA) at the dual frequencies of 1.6~GHz and 5~GHz in August, 2013. 
We detected a
single radio component at 1.6~GHz but nothing at 5~GHz, which suggests that the radio emission has a steep spectral 
index (S$_\nu\sim\nu^{\alpha}$) and $\alpha~<~-1.5$. The radio emission detected at 1.6~GHz has a size of 
7.5$\times$5~milliarcseconds or 8$\times$6~pc at a galactic distance of 250~Mpc. The emission has an integrated 
flux of 650~$\mu$Jy. Although it is possible that there 
are two SMBHs in the nucleus of KISS~1494 and only one has been detected in our observations, a more likely
interpretation of the radio emission is that it represents a coronal wind arising from the magnetized corona above 
the accretion disk or from the inner edge of the accretion disk or torus of the AGN. 

\subsection{2MASX~J1203}

2MASX~J1203 (or 2MASX~J12032061+1319316) also hosts a Seyfert~2 nucleus and shows double peaked emission lines in 
its SDSS spectrum (Figure~1) (for details see Rubinur et al. 2017), the only difference being that the peaks are not 
equal in this AGN. We carried out high resolution, radio observation of 2MASX~J1203 using the Karl G.~Jansky Very Large Array (EVLA) at 6 and 15~GHz in July, 2015 and May, 2016 respectively. We also used VLA archival data at 8.5 and 11.5~GHz,
from March, 2014 observations. We found that the radio emission has a prominent S-shaped morphology at all frequencies.
The S-shaped jets are very symmetric and extend out to a distance of 1.5$^{\prime\prime}$ (or 1.74~Kpc) on either side of 
a core of size 0.1$^{\prime\prime}$ (or 116~pc where the galaxy distance is assumed to be 245~Mpc) (Figure~2). The structure 
is similar to precessing radio jets observed at larger scale in galaxies (e.g. NGC~326) and our precession model gives a 
precession timescale of $\sim$10$^5$~years. We also obtained a similar age from spectral aging analysis using the spectral index maps.  We find that the precessing jets could be due to binary or dual SMBHs, a single SMBH with 
a tilted accretion disk or a dual SMBH in which a previous close passage of the SMBH caused the precession of the radio jets.

\begin{figure}[h]
\begin{minipage}{8cm}
\centering
\includegraphics[width=8.2cm]{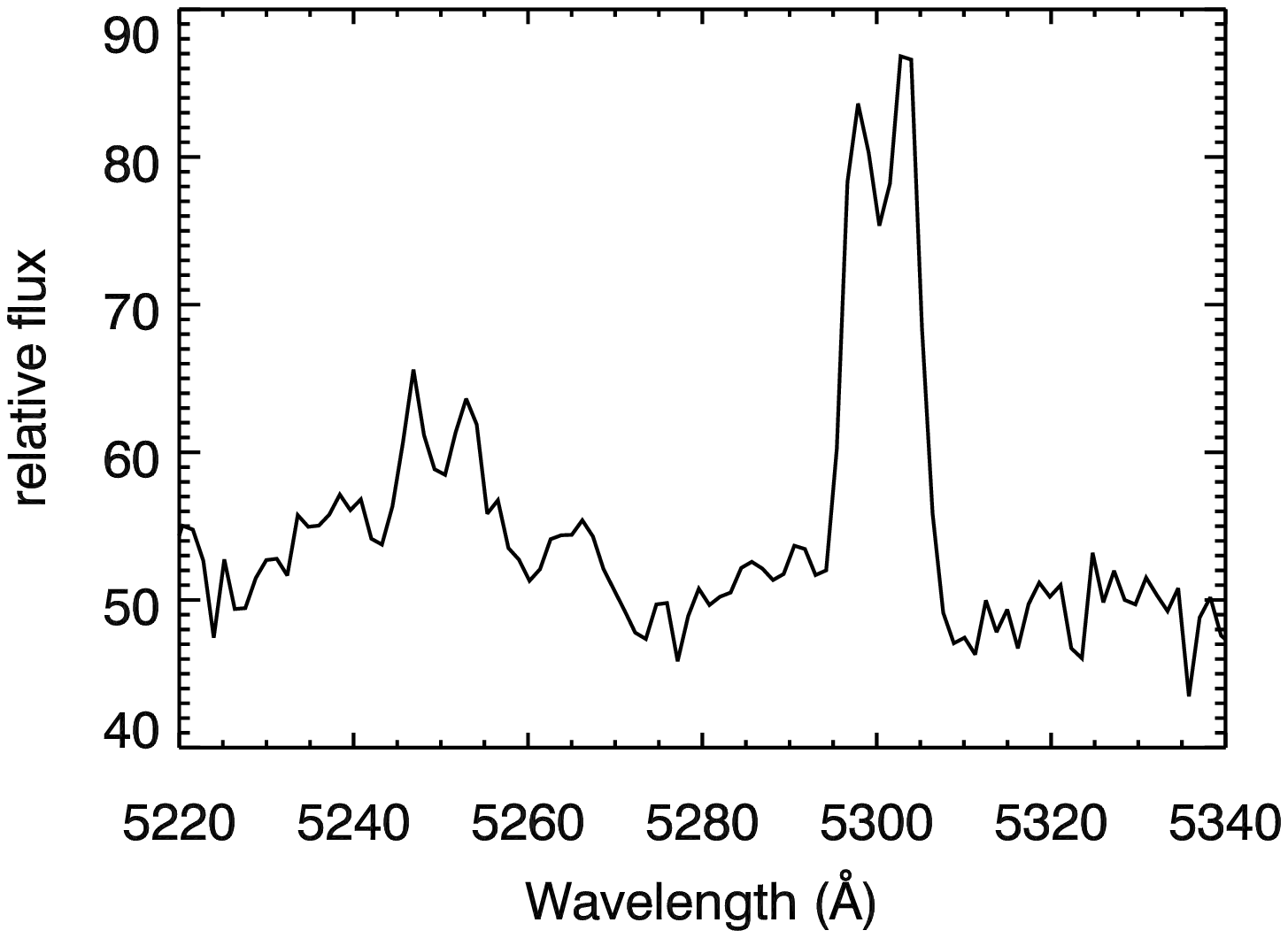}
\caption{This is the SDSS spectrum of the nucleus of 2MASX~J1203, showing the double peaked emission lines
in [OIII, 5007] and [OIII, 4959]. Note that the peaks are not equal, which suggests that the emission arises
from two different nuclei rather than a rotating disk.}
\end{minipage}
\hspace{2mm}
\begin{minipage}{8cm}
\centering
\includegraphics[width=9.4cm]{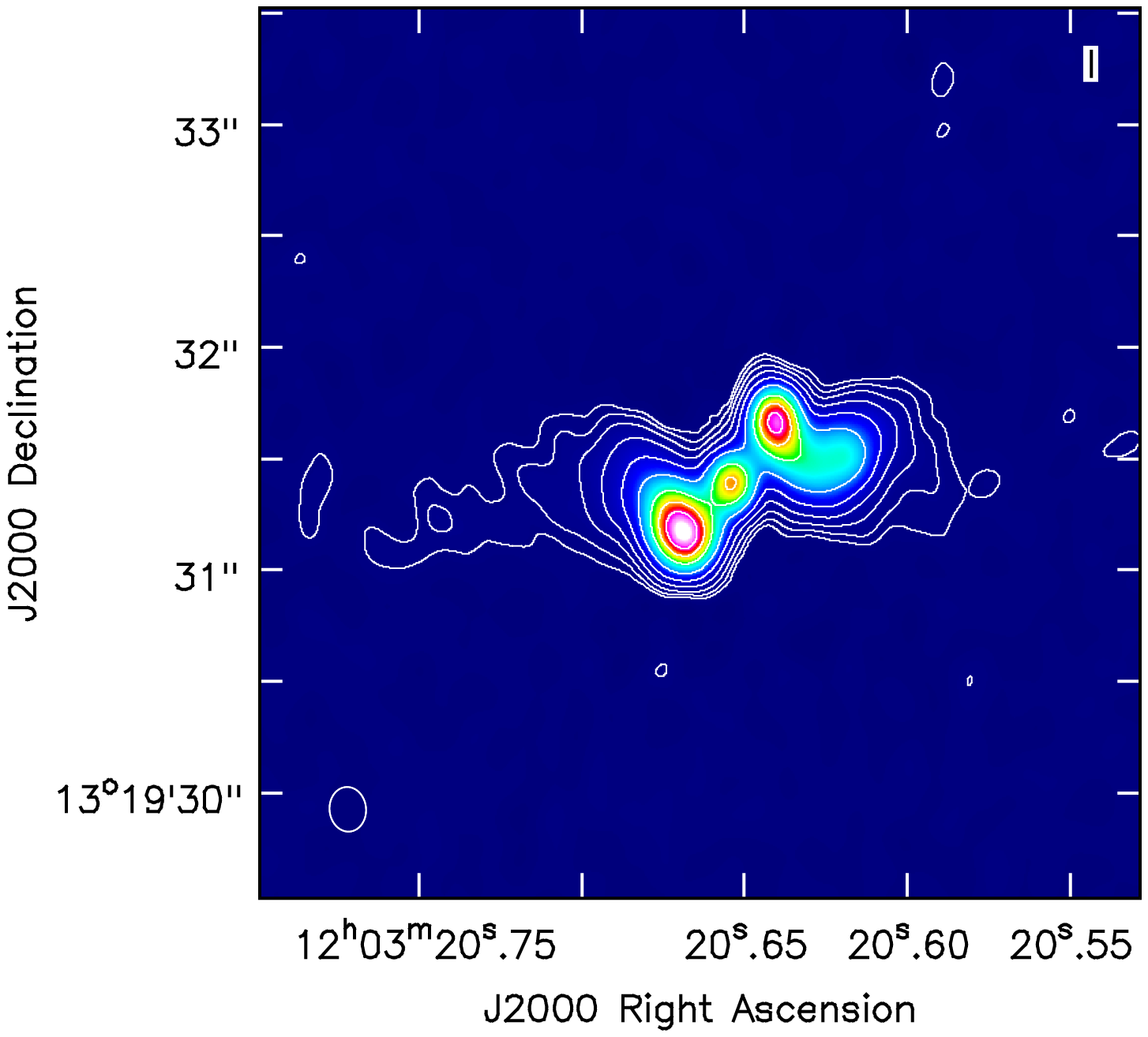}
\vspace{-6.5cm}
\caption{The above figure shows the 11.5~GHz EVLA image of 2MASX~J1203 made with with robust=0.5 (natural) weighting. 
The intensity contours are overlaid and have  0.0.60, 1.25, 2.5,
5, 10, 20, 40, 60, 80\% of the peak values of 5.5 mJy. The beam is $0.2^{\prime\prime}\times 0.16^{\prime\prime}$ (Rubinur et al. 2017).}
\end{minipage}
\end{figure}

\section{AGN Feedback and Star Formation Triggered by DAGN}

Dual or binary AGN will exert gravitational torques on the cold gas in the galaxy center, which will drive the gas into the very inner regions (Mayer et al. 2007). The pile-up of dense gas will result in nuclear star formation and often starburst 
activity. If the winds and outflows associated with the star formation/AGN are strong enough, the infalling gas will be 
driven away resulting in negative feedback. The negative feedback effects of DAGN may be larger than that of a single AGN. Furthermore, in DAGN we may also see  positive AGN feedback which arises due to the overlap and interaction of the outflows
due to the individual AGN. This can result in shocked gas that will cool and then fall back onto the molecular disk, thus producing positive  as well as negative AGN feedback. This can enhance the nuclear star formation, feed AGN accretion and perhaps help the two SMBHs coalesce faster. A good example of this scenario is NGC~6240 (Scoville et al. 2015) and MRK~266
(Mazzarella et al. 2012), where the extended outflows drive gas out of the nucleus (negative feedback) but dense gas has piled up between the nuclei and around the nuclei as a result of gas shocked by colliding gas flows. Some of the cool gas maybe due to shocked, cooled gas falling back onto the disk as well. 

\begin{figure}[h]
\begin{minipage}{16cm}
\centering
\includegraphics[width=14cm]{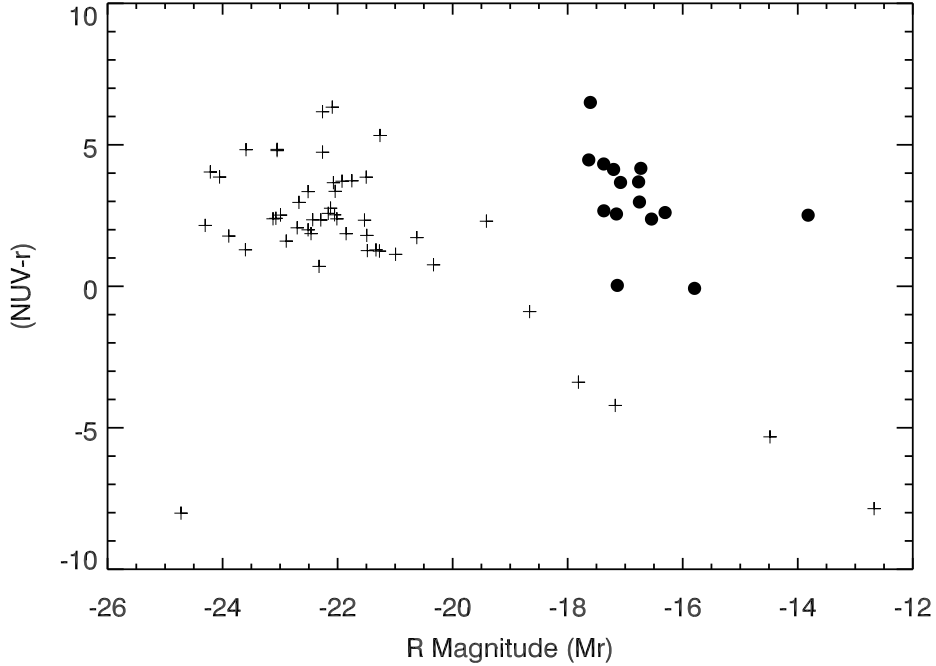}
\caption{The above figure is the (NUV-r) color plotted against the r band absolute magnitude of the 
confirmed DAGN (Table~1) and a sample of dual nuclei galaxies from Mezcua et al. (2014). The confirmed DAGN are filled black circles and the dual nuclei are marked with crosses.}
\end{minipage}
\end{figure}

To understand the effects of DAGN on star formation we derived the star formation rate from the near-UV (NUV) emission detected by GALEX from DAGN systems. Our results are shown in Table~2. We have assumed Kroupa stellar IMF, with constant 
star formation over 100~Myr. The star formation rate (SFR) is given by, 
SFR(UV)($M_{\odot}~yr^{-1}$~=~$3\times10^{-47}\lambda~L(\lambda)$, where $\lambda$ is the wavelength in Angstrom and L is the luminosity in erg/s. However, the NUV emission does contain a contribution from the AGN activity, but studies show that it varies from
8 to 20\% of the total NUV flux (Fujita et al. 2003). Of the 20 DAGN in Table~2, only two have SFR$>$10~$M_{\odot}~yr^{-1}$, of which one source is a 
high redshift binary quasar (LBQS 0103-2753) and for the other source, very little is known (SDSS J142607.71+353351.3).
Thus we find that DAGN do not generally have elevated SFRs; this suggests that the negative AGN feedback effects may be strong 
in these systems by driving away the gas.  

We also compared the SFR in DAGN with normal dual nuclei host galaxies that do not have AGN activity in both nuclei. The dual nuclei sample was obtained from Mezcua et al. (2014). We have derived the (NUV-r) color for both 
DAGN (Table~1) and dual nuclei galaxies. The color (NUV-r) is a measure of the star formation 
rate per unit stellar mass for these galaxies. We have plotted (NUV-r) against the absolute r band magnitude which is a 
measure of the stellar mass in these galaxies (Figure~3). We can immediately see that the (NUV-r) colors of both DAGN and dual nuclei galaxies are similar. Hence the presence of two AGN does not necessarily enhance the SFR in a galaxy.  We also 
obtain another interesting result, we find that the sources roughly divide into two groups according to r magnitude. 
Both groups have similar (NUV-r) colors but the confirmed DAGN appear to lie in low mass galaxies compared to the dual nuclei  that are in more massive (or red) galaxies. There are a few dual nuclei sources falling into the confirmed DAGN group; they may harbour DAGN that have not yet been detected. It is not clear whether the presence of two groups is an evolutionary effect, where the DAGN reside in less massive  galaxies evolving into more massive galaxies. We need to obtain a larger sample to confirm this.  

\section{Conclusions}

\noindent
{\bf 1.~}There are currently 26 confirmed DAGN described in the literature. DPAGN that can be identified from large spectroscopic surveys provide a reasonable method  of selecting a sample of DAGN candidates. DPAGN with higher stellar velocity dispersions may have a greater chance of harboring DAGN.\\
{\bf 2.~}We have done follow-up radio observations of two candidate DAGN that have double peaked emission 
lines in their optical spectra. We did not detect DAGN, but their presence cannot be ruled out in both 
cases. \\
{\bf 3.~}We have estimated the SFRs from the NUV fluxes of 20 DAGN galaxies. We find that the SFRs are 
generally $<~1~M_{\odot}~yr^{-1}$ and are not exceptionally elevated. We compared the (NUV-r) colors of DAGN 
host galaxies with galaxy mergers that have two nuclei but are not DAGN systems. The SFRs are similar for both 
groups. This suggests that it is the merging process and not the DAGN that contributes to the star formation in 
DAGN host galaxies.\\
{\bf 4.~}We find that in the (NUV-r) vs $M_{r}$ plot, the dual AGN all have lower absolute $M_{r}$ values compared to 
dual nuclei host galaxies. This suggests that dual AGN are more common in lower mass star forming galaxies.

%
%
\section*{Acknowledgements}
We acknowledge IIA for providing the computational facilities. 
The National Radio Astronomy Observatory is a facility of the 
National Science Foundation operated under cooperative agreement
by Associated Universities, Inc. This research has made use of the
NASA/IPAC Extragalactic Database (NED), which is operated by
the Jet Propulsion Laboratory, California Institute of Technology,
under contract with the National Aeronautics and Space Administration. 
Funding for the Sloan Digital Sky Survey IV has been
provided by the Alfred P. Sloan Foundation, the U.S. Department of
Energy Office of Science and the Participating Institutions. SDSSIV 
acknowledges support and resources from the Center for High Performance 
Computing at the University of Utah.

%
%
%

\footnotesize
\beginrefer

\refer Abbott, B. P.; Abbott, R.; Abbott, T. D.; et al. 2016, PhRvL, 116, 1102\\
\refer Barrows, R. Scott; Stern, Daniel; Madsen, Kristin; Harrison, Fiona; Assef, Roberto J.; Comerford, Julia M.; Cushing, Michael C.; et al. 2012, ApJ, 744, 7\\
\refer Begelman M. C., Blandford R. D., Rees M. J., 1980, Nature, 287, 307\\
\refer Bianchi, S.; Chiaberge, M.; Piconcelli, E.; Guainazzi, M.; Matt, G. 2008, MNRAS, 386, 105\\
\refer Burke-Spolaor S., Brazier A., Chatterjee S., Comerford J., Cordes J., Lazio
T. J. W., Liu X., et al. 2014, (arXiv:1402.0548)\\
\refer Civano, F.; Elvis, M.; Lanzuisi, G.; Jahnke, K.; Zamorani, G.; Blecha, L.; Bongiorno, A. et al. 2010, ApJ, 717, 209\\
\refer Calzetti, D. 2013, Secular Evolution of Galaxies, eds J.Falcón-Barroso and J.H.Knapen, Cambridge University Press\\
\refer Comerford, Julia M.; Schluns, Kyle; Greene, Jenny E.; Cool, Richard J. 2013, ApJ, 777, 64\\
\refer Comerford, J.M.; Pooley, D.; Barrows, R.S.; Greene, J.E.; Zakamska, N.L.; Madejski, G.M.; Cooper, M.C. 2015, ApJ, 806, 219\\
\refer Dasyra, K. M.; Combes, F.; Novak, G. S.; Bremer, M.; Spinoglio, L.; Pereira Santaella, M.; 
Salomé, P.; et al. 2014, A\&A, 565, 46\\
\refer 	Fabbiano, G.; Wang, Junfeng; Elvis, M.; Risaliti, G.  2011, Natur, 477, 431\\
\refer Fu, H.; Zhang, Z.; Assef, R.J.; Stockton, A.; Myers, A.D.; Yan, L.; Djorgovski, S. G.; et al. 2011, ApJ, 740L, 44\\
\refer Fujita, S.S.; Ajiki, M.; Shioya, Y.; Nagao, T.; Murayama, T.; Taniguchi, Y.; et al. 2003, ApJ, 586L, 115\\
\refer Garcia-Burillo, S.; Combes, F.; Usero, A.; Aalto, S.; Colina, L.; Alonso-Herrero, A.; Hunt, L. K.;
et al. 2015, A\&A, 580, 35\\
\refer Hayasaki, Kimitake; Mineshige, Shin; Ho, Luis C. 2008, ApJ, 682, 1134\\
\refer Hodges-Kluck, Edmund J.; Reynolds, Christopher S.  2012, ApJ, 746, 167\\
\refer Hudson, D. S.; Reiprich, T. H.; Clarke, T. E.; Sarazin, C. L.  2006, A\&A, 453, 433\\
\refer Junkkarinen, V.; Shields, G. A.; Beaver, E. A.; Burbidge, E. M.; Cohen, R. D.; Hamann, F.; Lyons, R. W. 2001, ApJ, 549L, 155\\
\refer Kelley, Luke Zoltan; Blecha, Laura; Hernquist, Lars  2017, MNRAS, 464, 3131\\
\refer Kharb, P.; Das, M.; Paragi, Z.; Subramanian, S.; Chitta, L. P.  2015, ApJ, 799, 161\\
\refer Komossa, S.; Burwitz, V.; Hasinger, G.; Predehl, P.; Kaastra, J. S.; Ikebe, Y.  2003, ApJ, 582L, 15\\
\refer Koss M., Mushotzky R., Treister E., Veilleux S., Vasudevan R., Trippe M. 2012, ApJ, 746, L22\\
\refer 	Leahy, J. P.; Williams, A. G.  1984, MNRAS, 210, 929L\\
\refer Lehto, Harry J.; Valtonen, Mauri J.  1996, ApJ, 460, 207L\\
\refer Liu, Xin; Civano, Francesca; Shen, Yue; Green, Paul; Greene, Jenny E.; Strauss, Michael A. 2013, ApJ, 762, 110\\
\refer Liu, Xin; Greene, Jenny E.; Shen, Yue; Strauss, Michael A. 2010, ApJ, 715, L30\\
\refer Mayer L., Kazantzidis S., Madau P., Colpi M., Quinn T., Wadsley J., 2007,
Science, 316, 1874\\
\refer Mazzarella, J. M.; Iwasawa, K.; Vavilkin, T.; Armus, L.; Kim, D.-C.; Bothun, G.; Evans, A. S.;
et al. 2012, AJ, 144, 125\\
\refer McGurk, R. C.; Max, C. E.; Medling, A. M.; Shields, G. A.; Comerford, J. M.  2015, ApJ, 811, 14\\
\refer McGurk, R. C.; Max, C. E.; Rosario, D. J.; Shields, G. A.; Smith, K. L.;
Wright, S. A. 2011, ApJ, 738L, 2\\
\refer Mezcua, M.; Lobanov, A. P.; Mediavilla, E.; Karouzos, M. 2014, ApJ, 784, 16\\
\refer Merritt, David; Milosavljevic, Milos 2005, LRR, 8, 8\\
\refer Murgia, M.; Parma, P.; de Ruiter, H. R.; Bondi, M.; Ekers, R. D.; Fanti, R.; Fomalont, E. B. 2001, A\&A, 380, 102 \\
\refer Muller-Sanchez, F.; Comerford, J. M.; Nevin, R.; Barrows, R. S.; Cooper, M. C.; Greene, J. E. 2015, ApJ, 813, 103\\
\refer Pringle, J. E. 1997, MNRAS, 292, 136\\
\refer Rees M. J., Begelman M. C., Blandford R. D., Phinney E. S., 1982, Nature,
295, 17\\
\refer Rodriguez, C.; Taylor, G. B.; Zavala, R. T.; Peck, A. B.; Pollack, L. K.; Romani, R. W. 2006, ApJ, 646, 49\\
\refer Rubinur, K.; Das, M.; Kharb, P.; Honey, M. 2017, MNRAS, 465, 4772\\
\refer Sakamoto, K.; Aalto, S.; Combes, F.; Evans, A.; Peck, A. 2014, ApJ, 797, 90\\
\refer Sanders, D. B.; Soifer, B. T.; Elias, J. H.; Madore, B. F.; Matthews, K.; Neugebauer, G.; Scoville, N. Z.  1988, ApJ, 325, 74\\
\refer Scoville, Nick; Sheth, Kartik; Walter, Fabian; et al. 2015, ApJ, 800, 70\\
\refer Silk, Joseph; Rees, Martin J. 1998, A\&A, 331L, 1\\
\refer Sillanpaa, A.; Haarala, S.; Valtonen, M. J.; Sundelius, B.; Byrd, G. G.  1988, ApJ, 325, 628\\
\refer Smith, K. L.; Shields, G. A.; Salviander, S.; Stevens, A. C.; Rosario, D. J. 2012, ApJ, 752, 63\\
\refer 	Tingay, S. J.; Wayth, R. B. 2011, AJ, 141, 174\\
\refer Valtonen, M. J.; Lehto, H. J.; Nilsson, K.; Heidt, J.; Takalo, L. O.; Sillanpaa, A.; Villforth, C.; et al. 2008, Natur, 452, 851\\
\refer Zhou, Hongyan; Wang, Tinggui; Zhang, Xueguang; Dong, Xiaobo; Li, Cheng 2004, ApJ, 604L, 33

\endrefer

\begin{table}
\caption{This table shows the NUV flux and the derived star formation rate of the sample of confirmed DAGN (Table~1). Four 
galaxies are omitted as they do not have GALEX NUV data}
\small
\begin{center}
\begin{tabular}{| l | r r r|}
\hline
Galaxy       & NUV Flux & Distance  & SFR \\
Name         & ($\mu Jy$)           & (Mpc)     & ($M_{\odot}~yr^{-1}$)\\
\hline
3c75             & 26.49 +/- 2.52  & 93.9       & 0.01426\\
NGC 326          & 29.03 +/- 5.03   &197        & 0.0688241\\
NGC 6240         & 931.27 +/- 9.78   &103       & 0.603547\\
MRK 739          & 764.39 +/- 16.81  &130       & 0.789156\\
NGC 3393         & 1401.31±20.75     &56.8      & 0.27618\\
CID 42           & 18.80±0.03        &126       & 0.0182\\
LBQS 0103-2753   & 114.39 +/- 3.64  &5247       & 192.3\\
SDSS J095207.62+25527.2 & 40.61 +/- 4.75 & 1740 & 7.51091\\
2MASX J11085103+0659014 & 50.73±5.38 & 853 & 2.25488\\
SDSS J1131-0204E & 18.02±3.99 & 667 & 0.489742\\
SDSS J133226.34+060627.3   & 8.46±1.89 & 986 & 0.502441\\
SDSS J142607.71+353351.3 & 0.35±0.21 & 7899 & 39.4499\\
SDSS J102325+324348 & 12.90  & 577 & 0.24449\\
SDSS J115822+323102 & 12.77 & 774 & 0.467341\\
SDSS J162345+080851 & 12.22 +/- 0.94 & 941 & 0.6610\\
SDSS 114642.47+511029.6 & 36.82 +/- 3.19 & 590 & 0.782977\\
MRK 266 &1224.72 +/- 23.28 & 119 & 1.059\\
ESO 509-IG066 NED 02 & 129.03 +/- 9.49 & 144 & 0.163447\\
NGC 3227 &4705.13+/- 48.16 & 20.4 &0.119617\\
SDSS J171544.02+600835.4 & 10.02 +/- 0.17 & 644.5 & 0.254258\\
\hline 
\end{tabular} 
\end{center} 
\end{table}

\end{document}